\documentstyle[11pt,appb_hepex,epsf]{article}   

\def\d0{D\O}
\def\D0{D\O}
\def\w{$W$}
\def\W{$W$}
\def\z{$Z$}
\def\Z{$Z$}
\def\wg{$W\gamma$}
\def\ww{$WW$}
\def\wz{$WZ$}
\def\wwg{$WW\gamma$}
\def\wwv{$WWV$}
\def\wwz{$WWZ$}

\def\pt{$p_T$}
\def\et{$E_T$}
\def\etmisv {\mbox{${\hbox{${\vec E}$\kern-0.6em\lower-.1ex\hbox{/}}}_T$}}
\def\etmis  {\mbox{${\hbox{$E$\kern-0.6em\lower-.1ex\hbox{/}}}_T$ }}

\def\ifmath#1{\relax\ifmmode #1\else $#1$\fi}%

\def\TeV{\ifmmode {\rm{ Te\kern -0.1em V}}\else
                   \rm{Te\kern -0.1em V}\fi}%
\def\GeV{\ifmmode {\rm{ Ge\kern -0.1em V}}\else
                   \rm{Ge\kern -0.1em V}\fi}%
\def\MeV{\ifmmode {\rm{ Me\kern -0.1em V}}\else
                   \rm{Me\kern -0.1em V}\fi}%
\def\GeVcc{\ifmmode {\rm{ \GeV/c^2}}\else
                   \rm{Ge\kern -0.1em V/c$^2$}\fi}%
\def\MeVcc{\ifmmode {\rm{ \MeV/c^2}}\else
                   \rm{Me\kern -0.1em V/c$^2$}\fi}%

\def\eg{{\it e.g.}}
\def\ie{{\it i.e.}}
\def\etal{{\it et al.}}

\begin{document}

\title{\vskip -3cm 
\rightline{Fermilab-Pub-97/079-E}
\rightline{D0-Conf-97-6}
\rightline{UCR/D0/97-07}
\vskip 2cm
W BOSON PHYSICS AT THE FERMILAB TEVATRON\thanks{Presented at
the Cracow Epiphany Conference on the \W\ Boson, 4-6 January 1997,
Krak\'{o}w, Poland.}  }

\author{JOHN ELLISON
\address{Physics Department,\\
University of California,\\
Riverside, CA 92521, USA}
}

\maketitle

\begin{abstract}
We present electroweak physics results from the \d0\ and CDF
experiments using data from $p \bar p$ collisions at $\sqrt{s} =
1.8$~\TeV.  Measurements of the cross sections times branching
ratios for \w\ and \z\ production, the inclusive width of the \w\
boson and the \w\ boson mass are presented.  Direct tests of the
\wwg\ and \wwz\ trilinear gauge boson couplings are also presented
based on studies of diboson production.

\end{abstract}

  
\section{Introduction}

The CDF and \d0\ detectors at the Fermilab Tevatron collider collected
data during the 1992-93 run (``run 1a'') corresponding to integrated
luminosities of $\approx$15~pb$^{-1}$ and $\approx$20~pb$^{-1}$ for
\d0\ and CDF respectively. In the 1994-95 run (``run 1b'') data sets
of $\approx$80~pb$^{-1}$ ({\d0}) and $\approx$90~pb$^{-1}$ (CDF)
were collected.
%
%
The large Tevatron data sets allow precise measurements of the \W\ boson
properties, such as its mass and width, and new studies of the physics
of electroweak boson pair production.

In this paper we present selected recent results on \w\ boson physics
from the analysis of a subset of these data, concentrating mainly on
electroweak physics. Theoretical aspects of \w\ production
at hadron colliders are covered in the talk by Stirling at this conference.

\section{W and Z Production and the Inclusive Width of the W Boson}

The measurement of the production cross-sections times branching
ratios ($\sigma \cdot B$) for the \W\ and \Z\ bosons allows a
determination of the width of the \W\ boson and a comparison of \W\
and \Z\ boson production with QCD predictions. The latter are
calculated to order $\alpha_s^2$ and therefore, precision measurements
of the cross sections are a test of radiative processes in QCD.

Measurements of the cross sections were made by \d0\ and CDF using the
$W~\to~e \nu$, $Z~\to~e e$, $W~\to~\mu \nu$ and $Z~\to~\mu \mu$ decay
channels.  Events were selected from single lepton triggers and
offline were required to contain a high \pt\ isolated lepton plus
missing transverse energy (\W\ events) or two high \pt\ isolated
leptons (\Z\ events). The \d0\ and CDF run 1a analyses are described
in~\cite{d0_xsec,cdf_xsec}.

\begin{figure}[ht]
    \epsfysize = 9.5cm
    \centerline{\epsffile{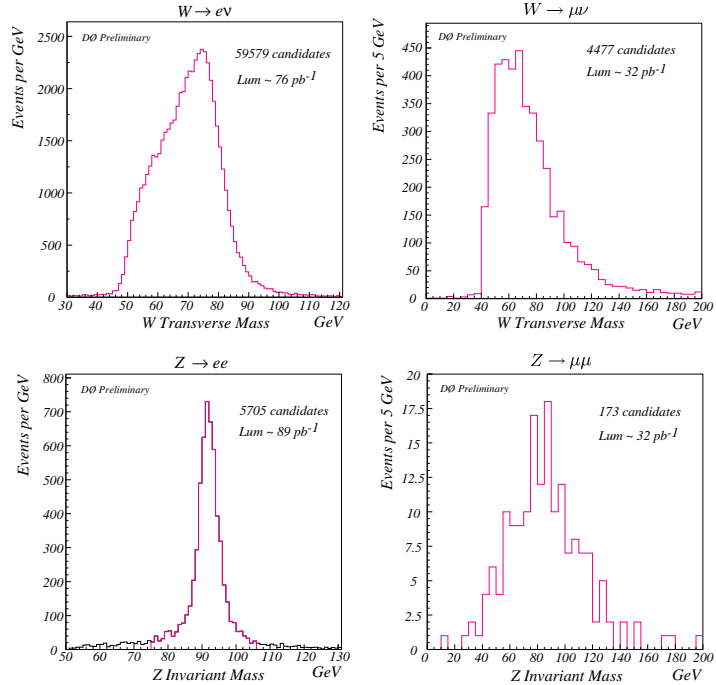}}
\caption{Transverse mass and invariant mass distributions for the \d0\ \W\
and \Z\ candidate events from run 1b.}
\label{fig:mtev-mee}
\end{figure}

Figure~\ref{fig:mtev-mee} shows the \d0\ preliminary run 1b transverse
mass spectra for $W~\to~\ell \nu$ candidate events and the invariant
mass spectra for $Z~\to~\ell \ell$ candidate events.

\begin{figure}[ht]
    \epsfysize = 5cm
    \centerline{\epsffile{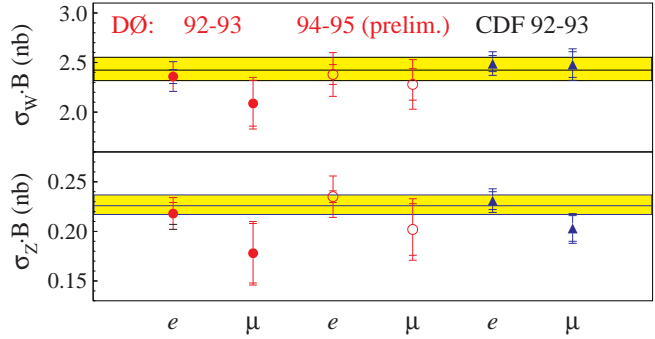}}
\caption{Measurements of the \W\ and \Z\ inclusive cross-sections
times branching ratios from \d0\ (circles) and CDF (triangles). Also
shown are the SM
theoretical predictions (shaded bands show central values and $\pm 1 \sigma$
uncertainties).}
\label{fig:xsec}
\end{figure}

The values of $\sigma \cdot B$ were calculated by subtracting the
background from the number of observed events and dividing by the
acceptance, efficiency and integrated luminosity. The results are
shown in Table~\ref{table:xsec} and are plotted in
Fig.~\ref{fig:xsec}.  The measurements agree well with the order
$\alpha_s^2$ QCD predictions (van~Neerven \etal~\cite{vanNeerven}) of
$\sigma_W \cdot B(W~\to~\ell \nu) =
2.42^{+0.13}_{-0.11}$~nb and $\sigma_Z \cdot B(W~\to~\ell \nu) =
0.226^{+0.011}_{-0.009}$~nb, shown as the shaded bands in
Fig.~\ref{fig:xsec}. The CTEQ2M parton distribution
functions~\cite{cteq}, and the values $m_W = 80.23 \pm 0.18$~\GeVcc\
and $m_Z = 91.188 \pm 0.002$~\GeVcc\ were used to calculate the
central values.
\begin{table}[ht]
\begin{center}
{\footnotesize 
\begin{tabular}{||l|c|c||c|c||} \hline\hline
                & \multicolumn{2}{c||}{ \D0  }
                & \multicolumn{2}{c||}{ CDF  }  \\ \hline
                & $e$ & $\mu$      
                & $e$ & $\mu$   \\ \hline
$W$ cand.          &  59579         & 4472 
                   &  13796         & 6222           \\
A$_W$ (\%)         & 43.4 $\pm$ 1.5 & 20.1 $\pm$ 0.7 
                   & 34.2 $\pm$ 0.8 & 16.3 $\pm$ 0.4 \\
$\epsilon_W$ (\%)  & 70.0 $\pm$ 1.2 & 24.7 $\pm$ 1.5 
                   & 72.0 $\pm$ 1.1 & 74.2 $\pm$ 2.7 \\
Bkg (\%)           &  8.1 $\pm$ 0.9 & 18.6 $\pm$ 2.0 
                   & 14.1 $\pm$ 1.3 & 15.1 $\pm$ 2.2 \\
$\int {\cal L}$ (pb$^{-1}$) 
                   & 75.9 $\pm$ 6.4 & 32.0 $\pm$ 2.7 
                   & 19.7 $\pm$ 0.7 & 18.0 $\pm$ 0.7 \\ \hline 
$Z$ cand.          &  5702          & 173 
                   &  1312          & 423            \\
A$_Z$ (\%)         & 34.2 $\pm$ 0.5 &  5.7 $\pm$ 0.5 
                   & 40.9 $\pm$ 0.5 & 15.9 $\pm$ 0.3 \\
$\epsilon_Z$ (\%)  & 75.9 $\pm$ 1.2 & 43.2 $\pm$ 3.0 
                   & 69.6 $\pm$ 1.7 & 74.7 $\pm$ 2.7 \\
Bkg (\%)           &  4.8 $\pm$ 0.5 &  8.0 $\pm$ 2.1 
                   &  1.6 $\pm$ 0.7 &  0.4 $\pm$ 0.2 \\
$\int {\cal L}$ (pb$^{-1}$) 
                   & 89.1 $\pm$ 7.5 & 32.0 $\pm$ 2.7 
                   & 19.7 $\pm$ 0.7 & 18.0 $\pm$ 0.7 \\

&&&&\\    \hline\hline
\end{tabular}
}
\end{center}
\caption[]{Summary of \d0\ and CDF analyses and production cross
section times branching ratio results for \W\ and \Z\ bosons. The
symbols A, $\epsilon$ and Bkg are the acceptance,
detection efficiency and background, respectively.}
\label{table:xsec}
\end{table}

The inclusive width of the \W\ boson is calculated using the measured
ratio ($R$) of the \W\ and \Z\ $\sigma \cdot B$ values:
$$R=\frac{\sigma_W \cdot B(W~\to~\ell \nu)}{\sigma_Z \cdot B(Z~\to~\ell \ell)}~\rm{~~with~~}~ B(W~\to~\ell \nu) = \frac{\Gamma(W~\to~\ell \nu)}{\Gamma(W)}$$

Many common sources of error cancel in $R$, including the uncertainty
in the integrated luminosity and parts of the errors in the acceptance
and event selection efficiency. The theoretical calculation $\sigma_W
/ \sigma_Z = 3.33 \pm 0.03$~\cite{vanNeerven} and the LEP measurement
$B(Z~\to~\ell \ell) = (3.367 \pm 0.0006)\%$~\cite{pdg} are used to
obtain $B(W~\to~\ell \nu)$. This measurement of $B(W~\to~\ell \nu)$
is then combined with a theoretical calculation~\cite{d0_xsec,Wwidth}
of $\Gamma(W~\to~\ell \nu) = 225.2 \pm 1.5$~\MeV\ to obtain an indirect
measurement of the \W\ inclusive width.

Figure 3 summarizes the measurements to date. The world average
(excluding the preliminary \d0\ run 1b results) is $\Gamma(W) = 2.062
\pm 0.059$~\GeV.  This can be compared with the Standard Model (SM)
prediction of $2.077 \pm 0.014$~\GeV~\cite{d0_xsec,Wwidth} to set
limits on non-standard decays of the \W. At the 95\% confidence level,
the upper limit on the width due to non-standard decays (\eg\ decays
to heavy quarks or to SUSY particles such as charginos and
neutralinos) is 109~\MeV\ (\ie\ approximately 5\% of the \w\ width).
\begin{figure}[ht]
    \epsfysize = 8cm
    \centerline{\epsffile{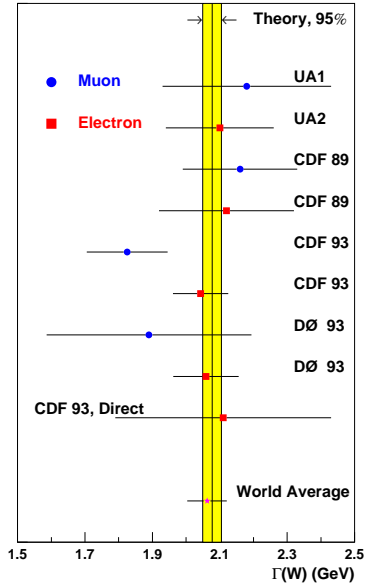}}
\caption{Measurements of the \W\ width from the Tevatron and CERN
compared with the SM theoretical prediction.}
\label{fig:width}
\end{figure}

The \w\ width directly affects the shape of the transverse mass
distribution in \w\ events. The effect is most prominent at high
values of $m_T$ where the Breit-Wigner line shape dominates over the
detector resolutions. CDF have utilized this fact to determine the \w\
width from a binned likelihood fit to the transverse mass distribution
in the region $m_T > 110~\GeVcc$~\cite{cdf_width_direct}. The result
is $\Gamma_W = 2.11 \pm 0.28 \pm 0.16~\GeV$. This method is currently
less precise than the indirect method described above but is
relatively independent of SM assumptions.

\section{Diboson Production and the Gauge Boson Couplings}

\begin{figure}[ht]
    \epsfysize = 4cm
    \centerline{\epsffile{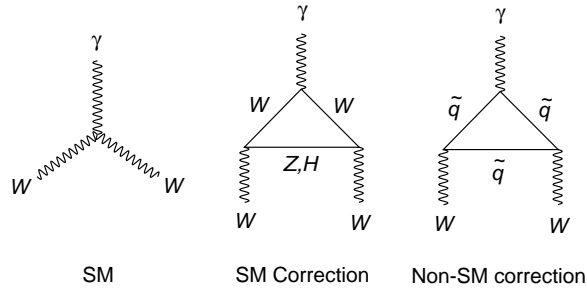}}
\caption{Standard Model \wwg\ vertex (left) and examples of loop corrections 
leading to deviations from the tree-level SM vertex functions.}
\label{fig:wwgvertex}
\end{figure}

The production of electroweak boson pairs at hadron colliders is
particularly interesting since these processes probe the nature of the
non-Abelian gauge boson self-couplings of the Standard Model (\ie\ the
\wwg\ and \wwz\ couplings). These processes are also sensitive to
deviations from the tree level SM couplings, which may arise due to
compositeness of the \W\ and \Z\ bosons or radiative loop corrections
to the vertex factors as shown in the examples of
Fig.~\ref{fig:wwgvertex}.
%
To test the agreement with the SM and to evaluate the sensitivity to
anomalous couplings the \wwv\ $(V=\gamma,Z)$ vertex is parametrized
using the phenomenological effective Lagrangian of
reference~\cite{wwvLagrangian}. Assuming electromagnetic gauge invariance, and
invariance under Lorentz and CP transformations the effective
Lagrangian is reduced to a function of five dimensionless coupling
parameters $g_1^Z, \kappa_V$, and $\lambda_V$.
%
%
In the SM at
tree level $g_1^Z = 1, \kappa_V = 1$ and $\lambda_V = 0$.  Assuming
that the coupling parameters for the \wwg\ and \wwz\ vertices are
equal, only two parameters remain:

\centerline{$\kappa_\gamma = \kappa_Z = \kappa~({\rm or}~\Delta\kappa
\equiv \kappa - 1)$}
\centerline{$\lambda_\gamma = \lambda_Z = \lambda.$}
\noindent
The coupling parameters are related to the lowest order terms in a
multipole expansion of photon interactions with the \W\ boson, \eg\
the \W\ magnetic dipole moment is given by $\mu_W =
(e/2m_W)(1 + \kappa_\gamma + \lambda_\gamma)$.

All interaction Lagrangians with constant anomalous couplings violate
unitarity at high energy. In the SM delicate cancelations ensure that
unitarity is satisfied and anomalous couplings would destroy these
cancellations leading to violation of unitarity.
Therefore, all the coupling parameters must be
modified to include form factors, \eg\
$$\Delta\kappa(\hat s) = \frac{\Delta\kappa^0}{(1 + \hat s / \Lambda^2)^n}$$ 
\noindent
where $\Delta\kappa^0 =$~value of coupling parameter at $\hat s = 0$,
$\hat s =$~square of the invariant mass of the partonic subprocess, $n
= 2$ for a dipole form factor and $\Lambda =$~form factor scale (a
function of the scale of new physics).

\subsection{\wg\ Production and the \wwg\ Coupling}

The study of \wg\ production at the Tevatron can be used to probe the
\wwg\ coupling~\cite{Wg}. The $u$- and $t$-channel Feynman diagrams
for $p \bar p~\to~\ell \nu \gamma$ correspond to photon bremsstrahlung
from an initial state quark, whereas the $s$-channel diagram is
sensitive to the \wwg\ vertex. Events in which a photon is radiated
from the final state lepton from single \W\ decay also result in the
same $\ell \nu \gamma$ final state, but are suppressed by imposing a
cut on the photon-lepton angular separation.
Non-SM values of the \wwg\ coupling result
in an increase in the total cross-section and an enhancement of events
with high-\pt\ photons. This is the experimental signature which is
used by \d0~\cite{d0_wg} and CDF~\cite{cdf_wg} to test for anomalous
couplings.

In addition to the \w\ selection criteria, a high-\pt\ isolated photon
with $p_T > 10$~(7)~\GeV\ is required in the \d0\ (CDF) analyses,
separated from the lepton by $\Delta R_{\ell \nu} > 0.7$ units in $\eta
- \phi$ space.  Photons are detected in the pseudorapidity range
$|\eta| < 1.1$ for CDF and $|\eta| < 1.1$ or $1.5 < |\eta| < 2.5$ for
\d0.

\begin{figure}[ht]
    \epsfysize = 4cm
    \centerline{\epsffile{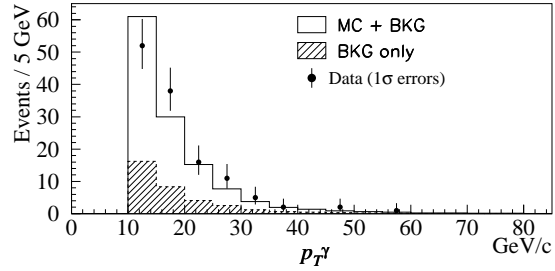}}
\caption{Distribution of photon transverse energy $p_T^\gamma$ for the
\d0\ combined run 1a plus 1b analyses. The shaded areas represent the
estimated background and the solid histograms are the expected signal
from the Standard Model plus the estimated background.}
\label{fig:wgpt}
\end{figure}

The numbers of signal events after background subtraction are compared
with the SM predictions in Table~\ref{table:wg}. Note that the number
of events is on the order of 100 for each experiment.  The \d0\
measured cross section times branching ratio (with $E_T^\gamma >
10~\GeV$ and $\Delta R_{\ell \nu} > 0.7)$ is $\sigma (W \gamma) \times
B(W~\to~\ell \nu) = 11.3^{+1.7}_{-1.5}~{\rm (stat)} \pm 1.5$~(sys)~pb
compared with the SM prediction of $\sigma (W \gamma) \times B(W \to
\ell \nu) = 12.5 \pm 1.0$~pb.  Figure~\ref{fig:wgpt} shows the \d0\
$p_T^\gamma$ distribution for the observed candidate events together
with the SM signal prediction plus the sum of the estimated
backgrounds. The number of observed events and the shapes of the
distributions show no deviations from the expectations. The SM
predictions for the number of events observed was obtained using the
leading order \wg\ event generator of Baur and Zeppenfeld~\cite{wgmc}
(using a K-factor of 1.34 to approximate higher order QCD effects)
combined with Monte Carlo simulations of the \d0\ and CDF detectors.
\begin{table*}[ht]
\begin{center}
{\footnotesize 
\begin{tabular}{||l|cc|cc||} \hline\hline
    & \multicolumn{2}{c|} { \D0 }
    & \multicolumn{2}{c||}{ CDF }  \\ 
    & \multicolumn{2}{c|} { 92.8 pb$^{-1}$ }
    & \multicolumn{2}{c||}{ 67.0 pb$^{-1}$ }  \\ \hline
    & $W\gamma\rightarrow e\nu\gamma $  
    & $W\gamma\rightarrow \mu\nu\gamma $
    & $W\gamma\rightarrow e\nu\gamma $  
    & $W\gamma\rightarrow \mu\nu\gamma $
\\ \hline
$N_{\rm data}$  & 57                      & 70 
                & 75                      & 34                  \\
$N_{\rm bkg}$   & 15.2 $\pm$ 2.5          & 27.7 $\pm$ 4.7 
                & 16.1 $\pm$ 2.4          & 10.3 $\pm$ 1.2      \\
$N_{\rm sig}$   & $41.8^{+8.8}_{-7.5}$    & $42.3^{+9.7}_{-8.3}$  
                & $58.9 \pm 9.0 \pm 2.6 $ & $23.7 \pm 5.9 \pm 1.1 $     \\
$N_{\rm SM}$    & 43.6 $\pm$ 3.1          & 38.2 $\pm$ 2.8 
                & $53.5 \pm 6.8$          & $21.8 \pm 4.3  $     \\
\hline\hline 
\end{tabular}
}
\end{center}
\caption{The number of candidate \wg\ events observed in the \d0\ and
CDF analyses ($N_{\rm data}$). $N_{\rm bkg}$ is the estimated
background and $N_{\rm sig}$ is the number of signal events after
background subtraction. Also shown is the SM prediction $N_{\rm SM}$.}
\label{table:wg}
\end{table*}

\begin{figure}[ht]
    \epsfysize = 4cm
    \centerline{\epsffile{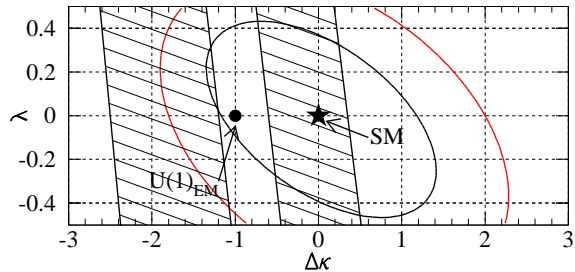}}
\caption[]{95\% CL limits on the \wwg\ couplings from \d0\ (inner
	ellipse) and CDF (outer ellipse). The shaded bands are the
	regions allowed by the CLEO 95\% CL limits from the
	observation of $b \to s \gamma$ decays~\cite{cleo}.}
\label{fig:wglimits}
\end{figure}

In both experiments, limits on the \wwg\ vertex coupling parameters
are obtained from a binned maximum likelihood fit to the photon \pt\
distribution.
In the electron channel
run 1b analysis, \d0\ require the electron-photon-neutrino transverse cluster
mass to be~$>90~\GeVcc$. This cut suppresses radiative \w\ decays and
increases the sensitivity to anomalous couplings by about 10\%.
Figure~\ref{fig:wglimits} shows the 95\% confidence level (CL) limits
in the $\Delta \kappa - \lambda$ plane, for a form factor scale of
$\Lambda = 1.5~\TeV$. Varying only one coupling at a time, the
following limits are obtained at the 95\% CL:
\begin{tabbing}
{\d0}:~~~~~~~~~~~~\=$-0.9 < \Delta \kappa < 0.9~~({\rm for}~\lambda = 0)$\\ 
		  \>$-0.3 < \lambda < 0.3~~({\rm for}~\Delta \kappa = 0)$\\
\> \\
CDF:              \>$-1.8 < \Delta \kappa  < 2.0~~({\rm for}~\lambda = 0)$\\
                  \>$-0.7 < \lambda < 0.6~~({\rm for}~\Delta \kappa  = 0)$
\end{tabbing}
The possibility of a minimal U(1)-only coupling ($\kappa =
\lambda = 0$) indicated by the star in Fig.~\ref{fig:wglimits} is
ruled out at the 88\% CL by the \d0\ measurement. This shows that the
photon couples not only to the \w\ electric charge but also to its
weak isospin.

\subsection{\ww\ and \wz\ Production}

\d0\ and CDF have searched for \ww\ and \wz\ production using two
decay modes: (a) the ``dilepton'' mode and (b) the ``lepton plus
jets'' mode.

Both experiments have searched for \ww\ production in the dilepton
decay modes $e \nu e \nu$, $e \nu \mu \nu$ and $\mu \nu \mu
\nu$~\cite{d0_wwdilep, cdf_wwdilep}. In the preliminary \d0\ run 1b
analysis, based on an integrated luminosity of 78.5~pb$^{-1}$, four
events pass the event selection criteria.
%
%
\d0\ set an upper limit on the cross section for $p \bar
p \to WW$ of 41~pb at the 95\% CL, estimated based on four candidate
events and a total estimated background of $2.6 \pm 0.4$ events.

In the CDF analysis based on 108~pb$^{-1}$ of data,
a SM \ww\ signal is observed above background. The event selection
yields 5 events with a background of $1.2 \pm 0.3$ events. The
measured \ww\ cross section is $\sigma (p \bar p \to WW) =
10.2^{+6.3}_{-5.1} \pm 1.6$~pb. The cross section for this process has
been calculated to next to leading order by Ohnemus~\cite{wwxs_nlo}
with the result $\sigma_{SM} (p \bar p \to WW) = 9.5 \pm 1.0$~pb.

The \w\ pair production process is sensitive to the \wwg\ and \wwz\
couplings, since the $s$-channel propagator can be a $\gamma$ or
\z. Anomalous couplings result in a higher cross section and an
enhancement of events with high \pt\ \w\ bosons. Due to the low
statistics in this channel, only the total cross section is used in
setting limits.  Assuming $\Delta \kappa_Z = \Delta
\kappa_\gamma$ and $\lambda_Z = \lambda_\gamma$, the limits obtained
on $\Delta \kappa$ and $\lambda$ are shown in Fig.~7.
%
\begin{figure}[ht]
\begin{center}
\begin{tabular}{c c}
    	\epsfxsize = 5.5cm
    	{\epsffile{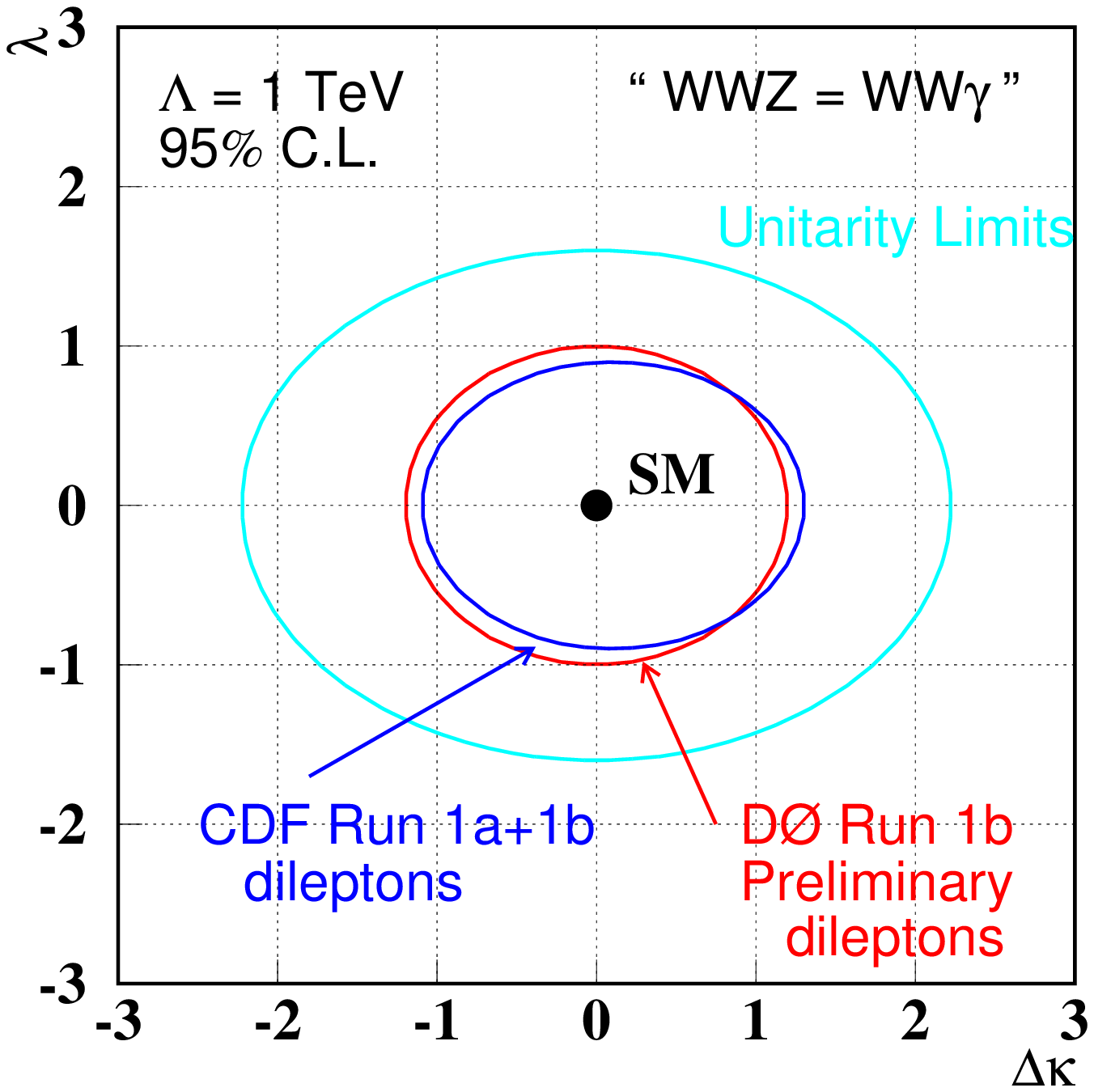}} &
    	\epsfxsize = 5.5cm
	{\epsffile{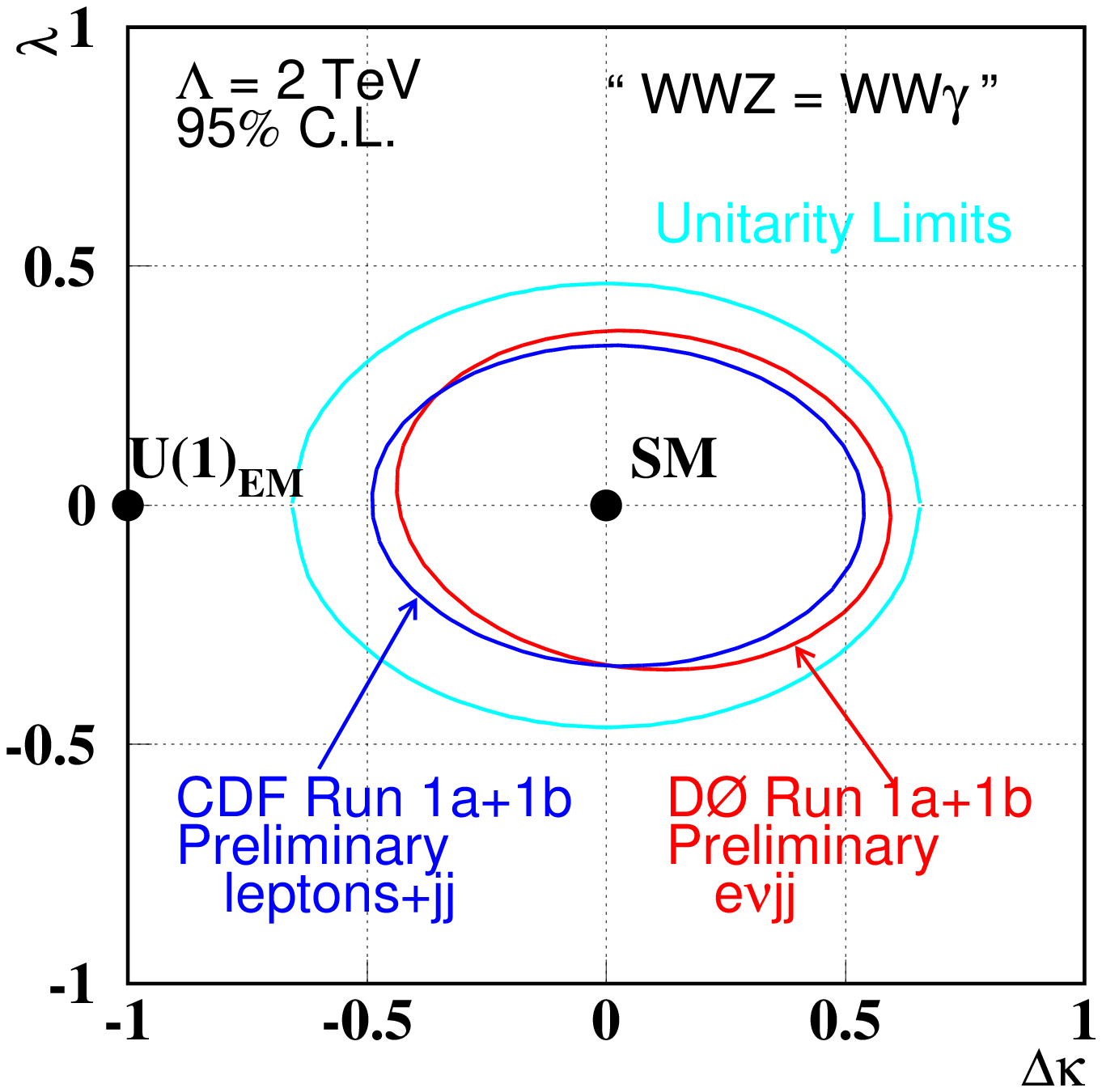}} \\
\end{tabular}
\caption{95\% CL limits from \d0\ and CDF in the $\Delta \kappa -
\lambda$ plane, assuming $g_1^Z = g_1^\gamma = 0$. Left: Limits from
$WW \to \ell \nu \ell^\prime \nu^\prime$. Right: Limits from the
$WW,WZ \to$ lepton(s) + $jj$ decay modes.}
\end{center}
\label{fig:wwvlim}
\end{figure}

In the lepton plus jets analyses \d0\ and CDF search for candidate
events containing a high-\pt\ lepton, missing \et\ and two jets with
invariant mass consistent with the \w\ or \z\ mass.  CDF also accept
events with two charged leptons and two jets resulting from $p \bar
p \to WZ \to \ell^+ \ell^- jj$.
The data are dominated by background, mainly from $W~+$~jets events
with $W \to e \nu$ and multijet production where one jet was
misidentified as an electron and there was significant (mismeasured)
missing \et. The data are in good agreement with the expectations.

At large values of $p_T^W$ the backgrounds are relatively small and it
is in this region where anomalous couplings would enhance the cross
section. CDF make a cut at $p_T^W > 200~\GeV$, where $p_T^W$ is
calculated from the \pt\ of the dijet system. Limits on anomalous
couplings are then derived by comparing the number of surviving events
with the predicted number of events as a function of the anomalous
coupling parameters, determined from 
the leading order calculation by Hagiwara, Woodside and 
Zeppenfeld~\cite{wwxs_lo} with a K-factor of 1.34.

\d0\ use a binned likelihood fit to the $p_T^W$ spectrum above $p_T^W
> 25~\GeV$, where $p_T^W$ is calculated from the \pt\ of the $e \nu$
system.  The results, assuming the \wwz\ and \wwg\ coupling parameters
are equal, are shown in Fig.~7.  At the 95\% CL for 
$\Lambda = 2.0~\TeV$, assuming
$\Delta \kappa_Z = \Delta \kappa_\gamma, \lambda_Z = \lambda_\gamma$
the following limits are obtained:
\begin{tabbing}
{\d0}:~~~~~~~~~~~~\=$-0.4 < \Delta \kappa  < 0.6~~({\rm for}~\lambda = 0)$\\
                  \>$-0.3 < \lambda < 0.4~~({\rm for}~\Delta \kappa  = 0)$\\
\> \\
CDF:              \>$-0.5 < \Delta \kappa  < 0.6~~({\rm for}~\lambda = 0)$\\
                  \>$-0.4 < \lambda < 0.3~~({\rm for}~\Delta \kappa  = 0)$
\end{tabbing}
Assuming SM \wwg\ couplings and fixing $\lambda_Z = 0$,
both experiments exclude the point $\kappa_Z = g_1^Z = 0$ at $>99\%$
CL. This is direct evidence for a nonzero \wwz\ coupling.

\section{Measurement of the W Boson Mass}

The mass of the \w\ boson $m_W$ is one of the fundamental parameters
of the SM. From the relation among the parameters of the SM we can
write $m_W$ in terms of $\alpha$, $G$, $m_Z$ and $\Delta r$. Radiative
corrections associated with the SM are contained in the $\Delta r$
term, the largest contributions to $\Delta r$ being proportional to
$m_t^2$ and ${\rm ln} m_H$. Therefore, a precise determination of the \w\
and top masses can be used to constrain the Higgs mass.

In $p \bar p \to W \to \ell \nu$ events at the Tevatron, the
longitudinal component of the neutrino cannot be determined and
therefore $m_W$ is measured using the transverse mass defined
by
\begin{displaymath} 
    m_T =  \sqrt{ 2\, p_T^\ell \, p_T^\nu \, 
                     (1 - \cos\varphi^{\ell\nu}) }
\end{displaymath} 
where $\varphi^{\ell\nu}$ is the angle between the lepton and neutrino
in the transverse plane.  The experimentally measured quantities are
the lepton momentum and the transverse momentum of the recoil system $
{\vec p}_T^{\,had}$. The transverse momentum of the neutrino is then
inferred from these two observables:
\begin{eqnarray*}
    \etmisv = - {\vec p}_T^{\,e} -  {\vec p}_T^{\,had} 
            = - {\vec p}_T^{\,e} - {\vec p}_T^{\,rec} - {\vec u}_T({\cal L})
\end{eqnarray*}
The \pt\ of the recoil system is given by
${\vec p}_T^{\,had} = {\vec p}_T^{\,rec} + {\vec u}_T({\cal L})$, 
where ${\vec p}_T^{\,rec}$ is the transverse momentum of the \w\ recoil 
and ${\vec u}_T({\cal L})$ is the transverse energy flow of the 
underlying event, which depends on the luminosity. The latter two
quantities are experimentally inseparable for \w\ events.

The \w\ mass is determined by generating Monte Carlo \w\ events and
fitting to the transverse mass distribution observed in the data.  To
correctly model the $m_T$ distribution a precise knowledge of the
response and resolution of the detector to charged leptons and the
recoil particles is required. The response determines the position of
the peak of the $m_T$ distribution, while the resolution determines
its width.

In CDF the momentum scale is determined by normalizing the measured
$J/\psi \to \mu^+ \mu^-$ mass peak to the world average value. The
calorimeter energy scale is then determined from a comparison
of the observed $E/p$ distribution with a detailed Monte Carlo
simulation. At \d0\ only $W \to e \nu$ events are used to measure
$m_W$. The electromagnetic calorimeter energy scale is
established using $Z \to e^+e^-$ events and calibrating against the
LEP measurement of $m_Z$, with additional constraints from
$J/\psi\to e^+e^-$ and $\pi^0 \to \gamma \gamma$ events.

In CDF the \w\ \pt\ is modeled in the Monte carlo using the \pt\
distribution of \z\ events observed in the data. The recoil system is
also taken from the \z\ data. The \d0\ experiment generates the \w\
\pt\ using the double differential \w\ production cross section in \pt\
and rapidity calculated to next to leading order by Ladinsky and
Yuan~\cite{Wprod_resum}. Minimum bias events are used to model the
underlying event. The response of the calorimeter to the hadronic
recoil is determined from the balance in transverse momentum in \z\
events.

Figure~8 shows the transverse mass distributions for
the data together with the best fit of the Monte Carlo
for the muon and electron channel run 1a data (CDF) and the run 1b
electron data (\d0).
\begin{figure}[ht]
\begin{center}
\begin{tabular}{c c}
    	\epsfysize = 5.5cm
    	{\epsffile{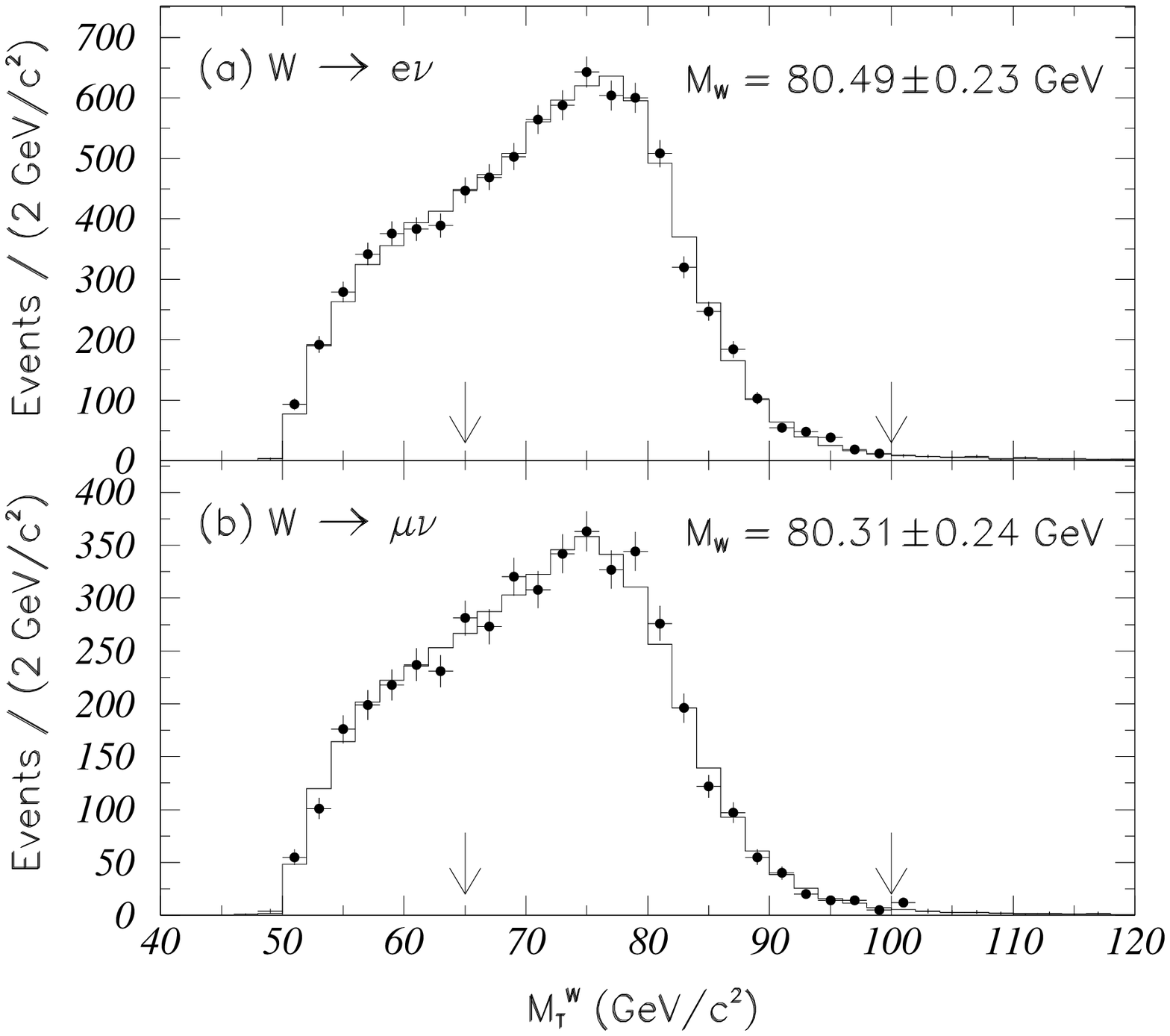}} &
    	\epsfxsize = 5.5cm
	{\epsffile{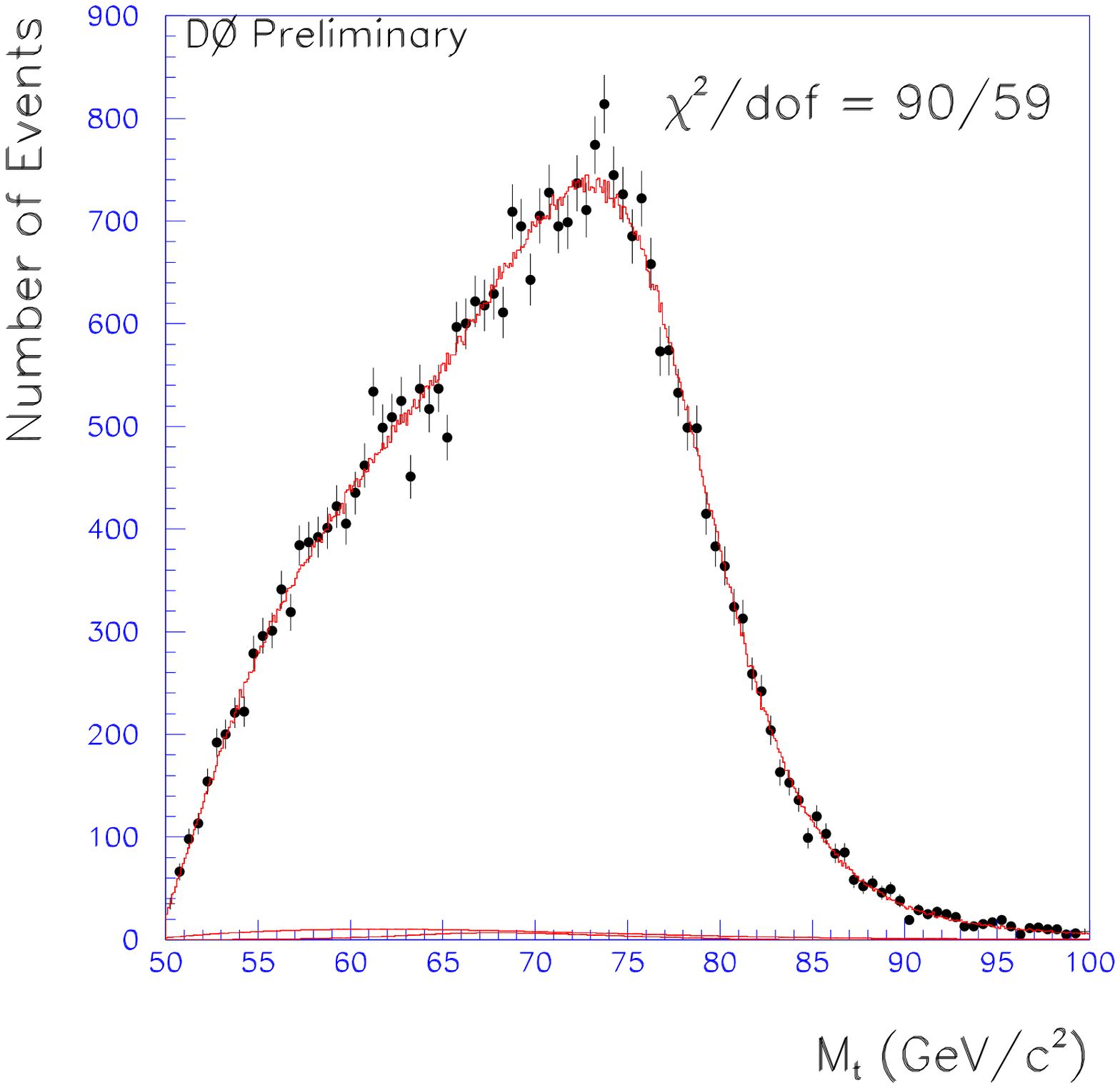}} \\
\end{tabular}
\caption{ Left: Transverse mass distributions from the CDF run 1a
data. The points are the data and the histogram is the best fit to the
data. The \w\ mass is obtained by fitting in the mass range indicated
by the arrows.  Right: Transverse mass distribution from the \d0 run
1b data.  The \w\ mass is obtained by fitting in the mass range $60 <
m_T < 90~\GeVcc$.}
\end{center}
\label{fig:mt_d0_cdf}
\end{figure}

The \w\ mass results are:
\begin{tabbing}
{\d0\ 1a}:~~ \=$m_W^e = 80.350 \pm 0.140 {\rm (stat)}
\pm 0.165 {\rm (syst)} \pm 0.160 {\rm (scale)}~\GeVcc$ \\

{\d0\ 1b}: \>$m_W^e = 80.380 \pm 0.070 {\rm (stat)}
\pm 0.130 {\rm (syst)} \pm 0.080 {\rm (scale)}~\GeVcc$\\

\end{tabbing}

\begin{tabbing}
{CDF 1a $W \to e \nu$}:~~~\=$m_W^e = 80.490 \pm 0.145 {\rm (stat)} \pm
0.175 {\rm (sys)}~\GeVcc$ \\

{CDF 1a $W \to \mu \nu$}: \>$m_W^\mu = 80.310 \pm 0.205 {\rm (stat)}
\pm 0.130 {\rm (sys)}~\GeVcc$ \\

\end{tabbing}
Table~\ref{table:mw_sys} lists the systematic errors on the individual
measurements and the common errors.
\begin{table*}[ht]
\begin{center}
{\footnotesize
\begin{tabular}{||l|rrr|rrr||} \hline\hline
    &   \multicolumn{3}{c|}{ CDF }
    &   \multicolumn{3}{c||}{ \D0 }                             \\ \hline
    &   \multicolumn{1}{c}{ e }
    &   \multicolumn{1}{c}{ $\mu$ }
    &   \multicolumn{1}{c|}{ common }
    &   \multicolumn{1}{c}{ 1a }
    &   \multicolumn{1}{c}{ 1b }
    &   \multicolumn{1}{c||}{ common }
\\ \hline
Statistical              & 145   & 205   &  ---  & 140 &  70 & ---    \\
Energy scale             & 120   &  50   &   50  & 160 &  80 &  25    \\
Angle scale              & ---   & ---   &  ---  &  50 &  40 &  40    \\
$E$ or $p$ resolution    & 80    &  60   &  ---  &  70 &  25 &  10    \\
$p_T^W$ and recoil model & 80    &  75   &   65  & 110 &  95 &        \\
PDF's                    & 50    &  50   &   50  &  65 &  65 &  65    \\
QCD/QED corr's           & 30    &  30   &   30  &  20 &  20 &  20    \\
$W$-width                & 20    &  20   &   20  &  20 &  10 &  10    \\
Backgrounds              & 10    &  25   &  ---  &  35 &  15 & ---    \\
Efficiencies             &  0    &  25   &  ---  &  30 &  25 & ---    \\
Fitting procedure        & 10    &  10   &  ---  &   5 &   5 & ---  \\ \hline
Total                    & 230   & 240   &   100 & 270 & 170 &  80    \\
\hline\hline
Combined                 & \multicolumn{3}{c|}  { 180 }
                         & \multicolumn{3}{c||} { 150}                \\
\hline\hline
\end{tabular}
}
\end{center}
\caption[]{Errors on $M_W$ in MeV/c$^2$. }
\label{table:mw_sys}
\end{table*}

\begin{figure}[ht]
    	\epsfysize = 8cm
    	\centerline{\epsffile{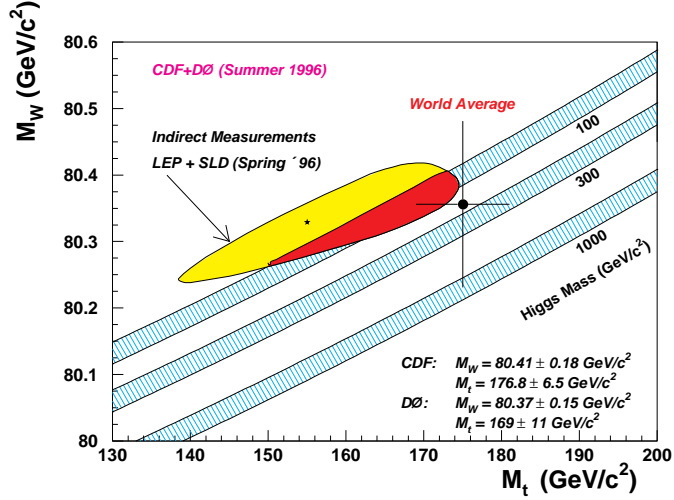}}
\caption{Direct measurements of $m_W$ and $m_t$ compared with the
predictions of the SM for different choices of Higgs boson mass. Also
shown is the $1 \sigma$ allowed region from fits to the LEP and SLC
data.}
\label{fig:mtmw}
\end{figure}

Using the above results and combining with previous \w\ mass
measurements~\cite{mw_previous}, a world average of $m_W = 80.356 \pm
0.125~\GeVcc$ is obtained.  The world average values of $m_W$ and
$m_t$ are shown in Fig.~\ref{fig:mtmw}. Also shown are the results
from the indirect measurements from LEP and SLC and the predictions of
the SM for various Higgs masses. The errors in the latter predictions
are indicated by the width of the bands and are primarily due to the
uncertainty in $\alpha(m_Z^2)$, the electromagnetic coupling strength
at the \z\ mass scale. Although only small improvements to the indirect
results from LEP/SLC data are expected, the direct measurements of
$m_W$ and $m_t$ are expected to improve by factors of $3-5$ from future
measurements~\cite{Tev2000} at the Tevatron. This would lead to a
prediction of the Higgs mass at the level of
$\delta m_H / m_H \approx 20\%$.

\subsection{Charge Asymmetry in $W \to \ell \nu$ Events}

On average $u$ valence quarks in the proton have higher momentum than
$d$ valence quarks. Therefore, $W^+$ bosons produced via $u \bar d \to
W^+$ are boosted predominantly along the proton direction and $W^-$
bosons from $\bar u d \to W^+$ are boosted predominantly along the
antiproton direction. This results in an asymmetry in the charged
lepton rapidity distribution in $W \to \ell \nu$ events, defined as
$$ A(y_\ell) =  
              { dN^+ (y_\ell) / dy_\ell \,-\, dN^- (y_\ell) / dy_\ell \over 
                dN^+ (y_\ell) / dy_\ell \,+\, dN^- (y_\ell) / dy_\ell } 
$$
where $N^{+(-)}$ is the number of positively (negatively) charged leptons 
with rapidity $y_\ell$. The $(1 \pm {\rm cos}\theta)^2$ 
decay asymmetry from the $V-A$ decay of
the \w\ must also be accounted for when studying $A(y_\ell)$.
The importance of this measurement is that it is sensitive to the
difference in the $u$ and $d$ quark distribution functions at very
high $Q^2$~($\approx m_W^2$) and low $x$ ($0.007 < x < 0.25$). The
measurements therefore constrain the parton distribution
functions (PDF's) and are used to reduce
the error in $m_W$ due to the uncertainty in the PDF's.

\begin{figure}[ht]
    \epsfysize = 7cm
    \centerline{\epsffile{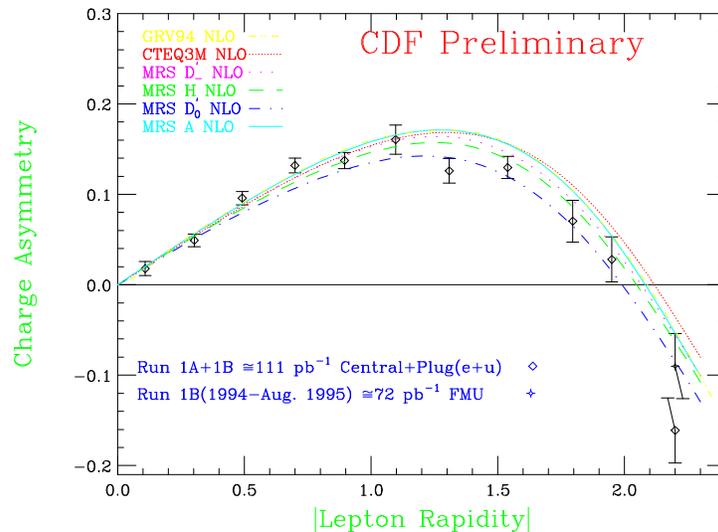}}    
\caption{CDF measurement of the lepton charge asymmetry
from  $W \to \ell \nu$ events compared with NLO 
predictions for different parton distribution functions. }
\label{fig:cdf_wasym} 
\end{figure}

The most recent measurements of this asymmetry from CDF are shown in
Figure~\ref{fig:cdf_wasym}. The data include the run 1a+1b results with a
total integrated luminosity of 111~pb$^{-1}$, utilizing the forward
muon toroids and the end plug electromagnetic calorimeter to
extend the coverage in rapidity. Also shown are the predictions of the
DYRAD NLO Monte Carlo~\cite{dyrad} using various parton
distribution functions.
The uncertainty in the measurement of $m_W$ due to the choice of PDF
in the modeling of \w\ events is constrained by
selecting only PDF's which show good agreement with the asymmetry
data,
\ie\ those for which $| \zeta | < 2$, where
\begin{equation} 
    \zeta = { \overline{A}_{\,PDF} \,-\, \overline{A}_{\,data} \over 
                  \sigma(\overline{A}_{\,data})  }
\end{equation}
and $\overline{A}$ is the weighted mean of the asymmetries $A(y_\ell)$.
The resulting uncertainty on $m_W$ is $50~\MeVcc$ in the CDF analysis
(see Table~\ref{table:mw_sys}).

\section{Conclusions}

Recent analyses of data based on the production of \w\ bosons at the Fermilab
Tevatron have resulted in precision measurement of the properies of the \w.
The indirect measurement of the \w\ boson width is 
$\Gamma(W) = 2.062 \pm 0.059$~\GeV\ (world average, excluding
the preliminary run 1b results), corresponding to a precision of 3\%.
The new world average value of the \w\ mass is  $m_W = 80.356 \pm
0.125~\GeVcc$, corresponding to a precision of 0.15\%.

New studies of the trilinear gauge boson couplings have also been
undertaken. The production of \wg\ and \ww\ events is in agreement with the
prediction of the SM, within the experimental sensitivity.
The current 95\% CL limits on the trilinear coupling parameters
are (approximately) $| \Delta \kappa | < 0.9 (0.5)$ and  
$| \lambda | < 0.3 (0.3)$ for the \wwg\ coupling assuming $\Lambda = 1.5~\TeV$
(\wwz\ coupling assuming $\Lambda = 2.0~\TeV$).
The measurements provide direct evidence for the existence of the
$WWZ$ coupling and show that the photon couples to the SU(2) weak isospin of
the \w\ boson as well as to its U(1) electric charge.

The upgraded \d0\ and CDF detectors will begin running at the Tevatron
in 1999. Improvements in the precision in the measurement of 
the properties of the \w\ boson, coupled with improved measurements of
the top quark mass, will yield predictions for the Higgs mass and may
provide crucial self consistency tests of the SM.
\vskip 7mm
\noindent
{\bf \centerline{Acknowledgments}}
\vskip 5mm
I would like to thank all my colleagues from the
\d0\ and CDF collaborations who helped in preparing my talk and this
paper. Particular thanks to Darien Wood for his useful comments.

\end{document}